\documentclass[%
 aip,
 jmp,%
 amsmath,amssymb,
 reprint,%
author-numerical,%
showkeys%
]{revtex4-1}

\usepackage{graphicx}

\usepackage{dcolumn}
\usepackage{bm}

\usepackage[english]{babel}

\usepackage{xcolor}

\definecolor{warm}{HTML}{ED1C25}
\definecolor{cold}{HTML}{21409A}
\definecolor{humid}{HTML}{018571}
\definecolor{dry}{HTML}{A6611A}

\begin{document}


\title{Condensation-induced flow structure modifications in turbulent channel flow\\%
	   investigated in direct numerical simulations}

\author{P. Bahavar}%
\email{philipp.bahavar@dlr.de}%
\affiliation{
German Aerospace Center (DLR), Bunsenstr.~10, 37073 G\"ottingen, Germany}
\author{C. Wagner}%
\affiliation{
German Aerospace Center (DLR), Bunsenstr.~10, 37073 G\"ottingen, Germany}
\affiliation{ Technische Universit\"at Ilmenau, Helmholtzring 1, 98693 Ilmenau, Germany
}%

\date{\today}

\begin{abstract}
The turbulent flow of a fluid carrying trace amounts of a condensable species through a differentially cooled vertical channel geometry is simulated using single-phase direct numerical simulations.
The release of latent heat during condensation is modeled by interdependent temperature and vapor concentration source terms governing the relation between the removal of excess vapor from the system and the associated local increase in fluid temperature.
A coupling between condensation and turbulence is implemented via solutal and thermal buoyancy.
When compared to simulations of an identical system without phase transition modeling, the modifications of the subcooled boundary layer due to the transient and highly localized release of latent heat could be observed.
A separate analysis of fluid before and after phase transition events shows a clear increase in post-interaction streak spacing, with the release of latent heat during condensation events opposing the cooling effect of the channel wall and the associated damping of turbulence.
\end{abstract}

\maketitle

%
\section{Introduction}\label{section:1-Introduction}
Condensation can occur when one or more condensable gas species are exposed to temperature differences around their dewpoint temperature.
In particular, environments connected to the atmosphere at standard ambient conditions are subject to the condensation of water vapor within the volume or at comparatively cool surfaces.
Studying the interaction between condensation and flow is therefore highly relevant to a wide range of applications, from meteorological flows to residential and vehicular ventilation.

Phase transitions in turbulent flows connect a variety of effects concerning both fluid dynamics and thermodynamics in the interplay between both phases.
While correlations and models can describe the heat transfer on a global level for specific fluid mixtures and parameter ranges\cite{Herranz1998, Bonner2013}, inclusion of phase transition in fully resolved simulations is possible as well.
The interaction can be realized via a two-way coupling, where a carrier phase determines the trajectories of the dispersed particles of the condensed phase, and the mass loading due to the particles in turn influences the flow of the carrier fluid. Another option is a four-way coupling, where particle-particle interactions are included in addition to the two-way coupling\cite{Kuerten2015}.
The particle motions can be calculated using Lagrangian methods \cite{Mashayek1998} or Eulerian models \cite{Fevrier2005,Masi2010}.
In these systems, the high computational costs associated with the simulation of the dispersed phase are unavoidable in order to capture the full dynamics of the flow.

In flows at ambient conditions in systems open to the atmosphere, the vapor concentration is typically small compared to the concentrations found in technical applications.
Additionally, the degree of subcooling is expected to be small and limited to regions in the immediate vicinity of cooled surfaces, where condensation therefore occurs predominantly.
Motivated by the expected low condensation mass flux, this study employs a single-phase approach to investigate the influence of the release of latent heat during condensation on the fundamental structure of turbulent flow in isolation from other phenomena of multiphase flow.

This approach extends the study by Russo \emph{et al.}\cite{Russo2013}, which compares differentially heated turbulent channel flow in three cases: with phase transition and droplets in a two-way coupled multiphase simulation, with inert particles, and without either phase transition or particles.
Here, a fourth configuration is investigated, where phase transition effects are included, but no droplets are simulated.

The condensable species is modeled by a scalar concentration field which is transported through the system by convection and diffusion.
The simulation can be considered as multiphase only in so far as the phase transition process is modeled, changing the density of the fluid as a result of the sudden change in temperature and concentration.
Consequently, the coupling with the flow is realized not by direct exchange of momentum but via the change in fluid density and the resulting buoyant forces.

This simplified approach allows the use of fully resolved direct numerical simulations (DNS) to simulate flows including phase transition effects while incurring only a small penalty in computational cost compared to DNS of mixed convection without phase transition.
The use of DNS allows an undisturbed evaluation of the simplified condensation modeling approach since the solution of the flow equations is free from the influence of turbulence or subgrid models, thereby isolating the effects caused by the release of latent heat at the phase boundary.

Building upon the well-understood system of turbulent flow through a differentially heated vertical channel geometry\cite{Kasagi1997}, which is extended to include an additional active scalar field representing the vapor concentration\cite{Bahavar2019}, this approach allows the analysis of modifications of turbulence due to the phase transition within the framework of turbulent flow in the presence of buoyancy gradients.
%
%
%
\section{Governing equations}\label{section:2-Governing_equations}
The flow of the gas mixture is simulated by directly solving the incompressible Navier-Stokes equations for the carrier phase,
\begin{align}
\nabla\cdot \mathbf{u}=0,\label{eq:continuity}\\
\frac{\partial\mathbf{u}}{\partial t} + (\mathbf{u}\cdot\nabla )\mathbf{u} = -\frac{1}{\rho}\nabla p + \nu \nabla^2\mathbf{u}-\mathbf{B},\label{eq:momentum}
\end{align}
where $\mathbf{u}=(u_x,u_y,u_z)$ is the vector field describing the fluid velocity as a function of space and time, $\nu$ and $\rho$ represent the kinematic viscosity and density of the fluid, and $\nabla p$ is the pressure gradient driving the flow.
The evolution of the temperature $T$ and vapor concentration $c$, expressed in terms of molar fraction, are modeled with convection-diffusion equations,
\begin{align}
&\frac{\partial T}{\partial t} + \mathbf{u}\cdot\nabla T = \kappa\nabla^2 T + \frac{h_v}{c_p} f(T,c),\label{eq:cd_temperature}\\
&\frac{\partial c}{\partial t} + \mathbf{u}\cdot\nabla c = D\nabla^2 c - f(T,c),\label{eq:cd_concentration}
\end{align}
with the thermal diffusivity $\kappa$ and the mass diffusivity $D$.
The inclusion of these scalar fields into the simulation gives rise to buoyant forces $\mathbf{B}$ due to density changes induced by temperature and concentration differences within the system.
The buoyancy contributions are linearized within the framework of the Boussinesq approximation\cite{Gray1976},
\begin{align}
\mathbf{B} = \beta_T (T-T_{\mathit{ref}})\mathbf{g} + \beta_c (c-c_{\mathit{ref}})\mathbf{g}.\label{eq:buoyancy}
\end{align}
Here $\mathbf{g}$ is the gravitational acceleration and $\beta_T$ and $\beta_c$ refer to the expansion coefficients with respect to temperature and vapor concentration \cite{Bird2015},
\begin{align}
\beta_{T}=\frac{1}{\rho_{\mathit{ref}}}\left.\frac{\partial\rho}{\partial T}\right\rvert_{T_{\mathit{ref}}},\quad\beta_{c}=\frac{1}{\rho_{\mathit{ref}}}\left.\frac{\partial\rho}{\partial c}\right\rvert_{c_{\mathit{ref}}}.\label{eq:expansion_coeffs}
\end{align}
$T_{\mathit{ref}}$ and $c_{\mathit{ref}}$ are set reference values about which the density variation is calculated\cite{Barletta1999}, and $\rho_{\mathit{ref}}=\rho(T_{\mathit{ref}},c_{\mathit{ref}})$ is the reference density.

The coupling between temperature and concentration caused by phase transition effects is represented by the source term $f$ in equation (\ref{eq:cd_concentration}), which describes the loss of vapor concentration due to condensation.
Conversely, the temperature source term in equation (\ref{eq:cd_temperature}) differs by its sign and the ratio of latent heat of condensation to specific heat capacity of the fluid, $h_v/c_p$, thereby yielding the increase in temperature associated with the release of latent heat at the phase transition.
The mass flux $\dot{m}^\ast$ across the phase boundary during condensation is given by the simplified Hertz-Knudsen-Schrage equation\cite{Marek2001},
\begin{align}
\dot{m}^\ast = \frac{2\sigma_c}{2-\sigma_c}\sqrt{\frac{M}{2\pi RT}}(p_v - p_{\mathit{sat}}),\label{eq:HKS_mass_flux}
\end{align}
with the molar weight of the vapor $M$ and universal gas constant $R$.
$\sigma_c$ is the condensation accommodation coefficient, describing the probability of a vapor molecule remaining in the liquid phase after impinging on the physical boundary between the phases.
$p_v$ and $p_{\mathit{sat}}$ are the partial vapor pressure and saturation pressure, respectively, related to the vapor concentration via $c=p_v/p_{\mathit{total}}$.
In terms of the concentration, the rate of change in vapor content $f$ in a volume $V$ across a boundary of area $A$ is then given by
\begin{align}
f = \frac{2\sigma_c}{2-\sigma_c}\sqrt{\frac{RT}{2\pi M}}\frac{A}{V}(c - c_{\mathit{sat}}).\label{eq:HKS_vapor_source}
\end{align}
Vapor removed from the system by the source term is not considered further.
The liquid phase is not simulated in this single-phase approach.

The interdependence of the scalar transport equations via the phase change source terms in addition to the coupling between the equations for the scalar fields and the velocity field via convective transport on the one hand and buoyancy on the other completes the full coupling between the governing equations.
%
%
\section{Validation}
The validity of the numerical scheme for solving the governing equations in a simulation including scalar transport and buoyant forces along the streamwise direction is tested in a generic biperiodic channel geometry.

The finite volume DNS is performed using \mbox{OpenFOAM}, with second-order central differencing in space and explicit second-order accurate leapfrog-Euler time integration\cite{Kath2016}.
The projection method is used to determine a pressure field which corrects the initial solution for the velocity field by removing divergences\cite{Chorin1968}.

To solve the governing equations, the channel geometry is discretized into $N_x \times N_y \times N_z = 396 \times 180 \times 316$ hexahedral cells.
These cells are uniformly distributed in the streamwise ($x$) and spanwise ($z$) direction, with a resolution of $\Delta x^+=6$ and $\Delta z^+=3$, given in wall units $\nu/u_{\tau}$.
Along the wall-normal ($y$) direction, the cells are distributed following a hyperbolic tangent function, increasing the number of cells near the walls, where high gradients need to be resolved, while limiting the number of cells located in the bulk to reduce computational costs and at the same time minimizing adverse effects due to cell-to-cell stretching\cite{Vinokur1983}.
The resulting spacing ranges from $\Delta y^+=0.2$ to $3.7$.

To establish the accuracy of the DNS solver, a channel without the vapor concentration field including only the temperature is simulated.
Isothermal boundaries are applied at the walls, with temperatures set to $T_h$ at the heated and $T_c$ at the cooled wall.
No-slip and impermeability boundary conditions are imposed on the velocity field.

The flow is characterized by a bulk Reynolds number
\begin{align}
\mathrm{Re} = \frac{u_b \delta}{\nu} = 2280\label{eq:Re_validation}
\end{align}
with the bulk flow velocity $u_b$ and channel half-width $\delta$.
Buoyant forces are quantified by the Grashof number
\begin{align}
\mathrm{Gr} = \frac{g\delta^3\beta_T\Delta T}{\nu^2} = 120000\label{eq:Grashof_thermal_validation}
\end{align}
with the characteristic temperature difference \mbox{$\Delta T = T_h - T_c$.}
This results in a flow which is primarily determined by forced convection with a Richardson number of
\begin{align}
\mathrm{Ri}=\frac{\mathrm{Gr}}{\mathrm{Re}^2}=0.023.\label{eq:Richardson_thermal_validation}
\end{align}
The temperature diffusion is given by the Prandtl number $\mathrm{Pr}=\kappa/\nu=0.71$, corresponding to dry air.

\begin{figure}[t]
\centering
\includegraphics[scale=1]{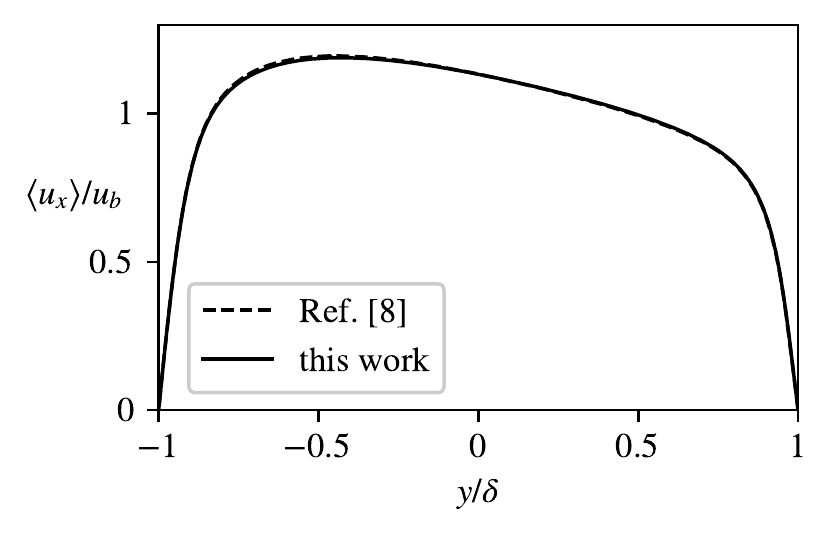}
\caption{Streamwise velocity profile obtained by DNS compared to reference data.}\label{fig:validation_ux}
\end{figure}
\begin{figure}[t]
\centering
\includegraphics[scale=1]{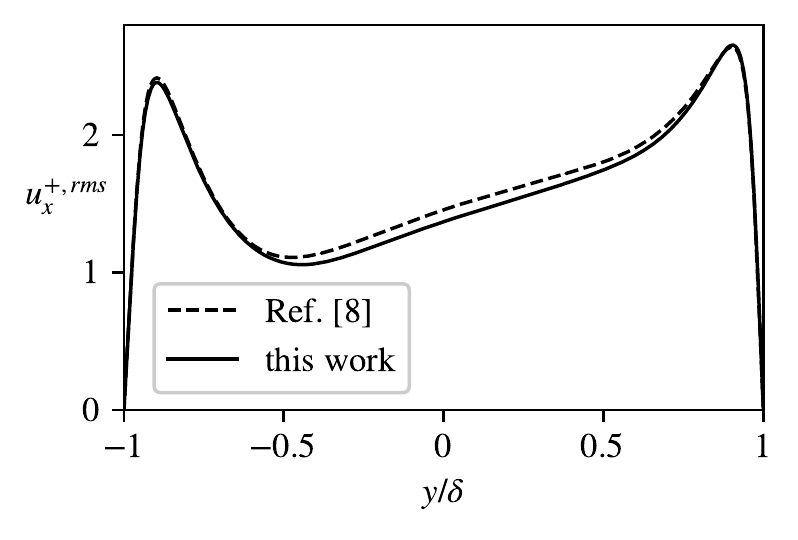}
\caption{Profile of streamwise velocity fluctuations normalized with $u_\tau$ compared to reference data.}\label{fig:validation_ux_rms}
\end{figure}
\begin{figure}[t]
\centering
\includegraphics[scale=1]{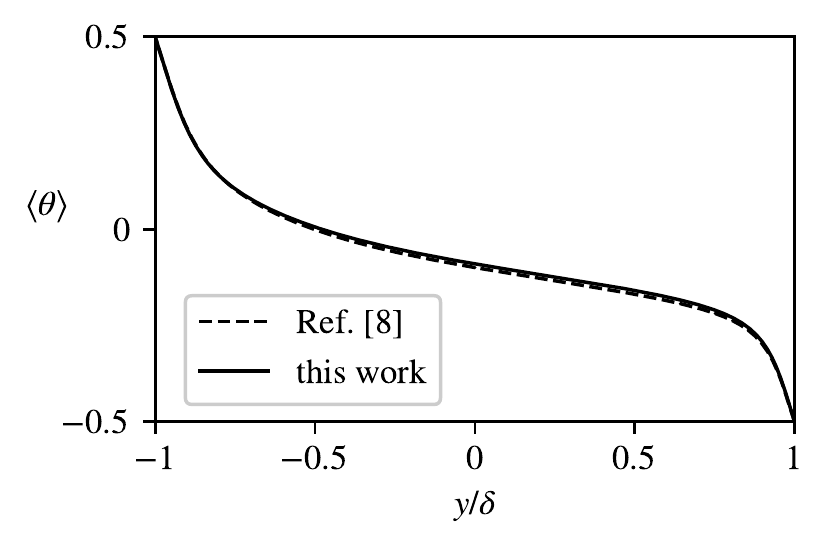}
\caption{Temperature profile obtained by DNS compared to reference data.}\label{fig:validation_T}
\end{figure}
\begin{figure}[t]
\centering
\vspace*{-1.6mm}
\includegraphics[scale=1]{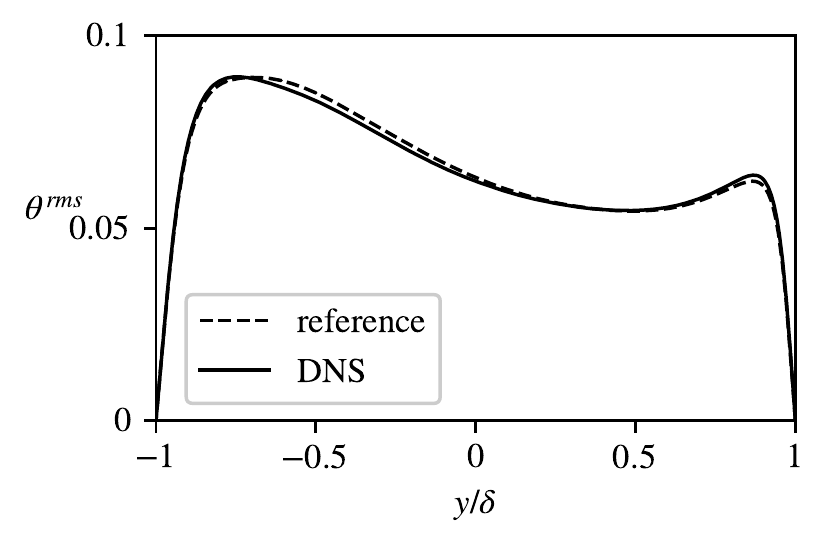}
\caption{Profile of temperature fluctuations obtained by DNS compared to reference data.}\label{fig:validation_T_rms}
\end{figure}

The flow parameters are chosen to allow comparison with an established reference DNS for differentially heated turbulent flow in a vertical channel geometry\cite{Kasagi1997}.
For statistical analysis, averaging is performed in time over an interval of $\Delta t^+ = 30$, with the dimensionless time unit $t^+=t u_{\tau}/\nu$, and along the homogeneous spatial directions.
Averaged quantities are denoted by $\langle\cdot\rangle$.

The nondimensional normalized temperature
\begin{align}
\theta = \frac{T-T_{\mathit{ref}}}{\Delta T}\label{eq:def_theta}
\end{align}
is introduced, with $T_{\mathit{ref}}=(T_h + T_c)/2$ used as the reference for this configuration.

Figures \ref{fig:validation_ux} and \ref{fig:validation_ux_rms} show the profile of the average streamwise velocity $u_x$ and the corresponding average fluctuations $u_x^{\mathit{rms}}$.
Very good agreement with the reference is found for the mean velocity at all points.
The fluctuations of the streamwise velocity are fully captured in the highly turbulent regions near the channel walls, but underestimated in the bulk region of the flow.

Analogously, Figures \ref{fig:validation_T} and \ref{fig:validation_T_rms} show the mean temperature and and its fluctuations.
Again, the agreement between the results obtained here and the reference is good for the mean field, with slight deficits in the magnitude of the fluctuations in the same regions as observed for the velocity.
This suggests that the error in the temperature fluctuations is a result of reduced mixing due to the underestimated turbulence intensity.
This discrepancy is expected due to the fundamental differences between the second-order finite volume approach employed in this study compared to the spectral method used for the reference case.

In addition to assessing solver accuracy, the biperiodic channel setup can be used to investigate the applicability of the Boussinesq approximation in the case of added solutal buoyancy.
To this end, uniform concentration boundary conditions are applied at the channel walls such that the buoyant forces caused by the resulting concentration gradients act in the same direction as the thermal buoyancy.
The solutal Grashof number is
\begin{align}
\mathrm{Gr}_c = \frac{g\delta^3\beta_c\Delta c}{\nu^2} = 248000\label{eq:Grashof_solutal_validation}
\end{align}
with $\Delta c$ describing the difference in concentration between the values at the walls.
The Richardson number quantifying the combined influence of solutal and thermal buoyancy is then $\mathrm{Ri}=0.07$.
The diffusive transport of the concentration field is quantified by the Schmidt number, $\mathrm{Sc}=D/\nu=0.48$.

In this configuration, the high vapor load required to cause high solutal buoyancy means that the assumption of constant fluid properties apart from the linearized density variation is not well justified\cite{Bahavar2019}.
Particularly, the coefficients for scalar transport are sensitive to the mixing ratio between carrier fluid and vapor, varying by more than 10\% across the range of $\Delta c$.
To investigate the inaccuracies introduced by applying the Boussinesq approximation regardless of these caveats, simulations were performed using an extended approximation including linearized changes of the thermal diffusivity $\kappa(c) = \beta_{c,\kappa}\,(c-c_{\mathit{ref}})$, analogously to the treatment of the density in equation (\ref{eq:buoyancy}).
The results obtained using this approach are compared to a simulation using the strict Boussinesq approximation, where the thermal diffusivity is fixed.

\begin{figure}[t]
\centering
\includegraphics[scale=1]{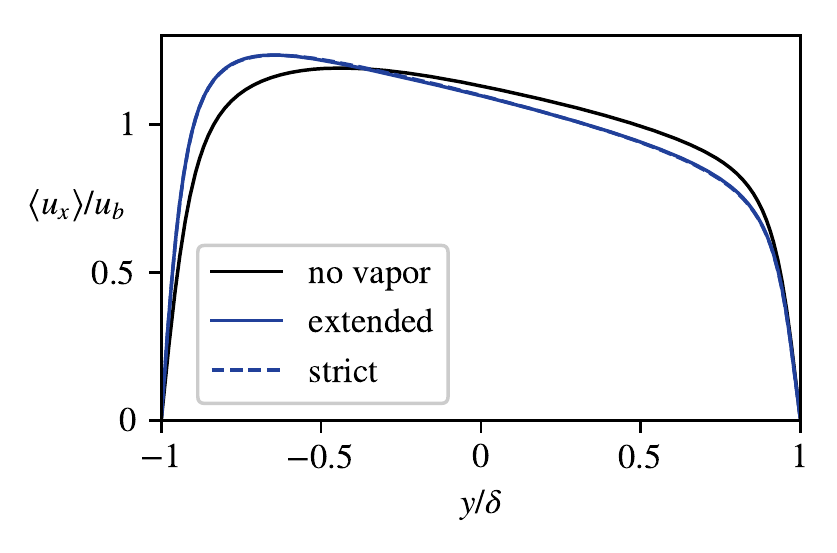}
\caption{Average streamwise velocity profile for channel flow with combined thermal and solutal buoyancy, comparison of strict and extended approximation.The black line shows the profile for the system without solutal buoyancy for comparison.}\label{fig:boussinesq_ux}
\end{figure}
\begin{figure}[t]
\centering
\includegraphics[scale=1]{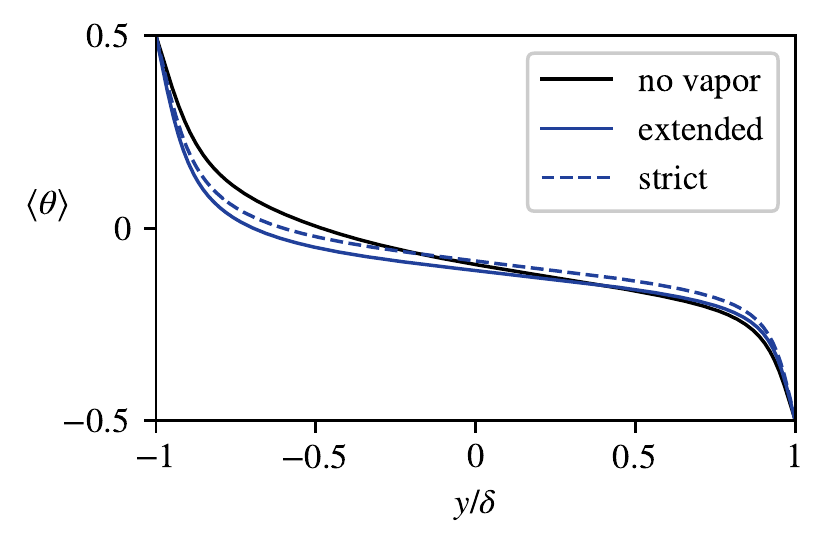}
\caption{Average temperature profile for channel flow with combined thermal and solutal buoyancy, comparison of strict and extended approximation.The black line shows the profile for the system without solutal buoyancy for comparison.}\label{fig:boussinesq_T}
\end{figure}

Figures \ref{fig:boussinesq_ux} and \ref{fig:boussinesq_T} show the resulting velocity and temperature profiles compared to the previous validation case without solutal buoyancy.
The effect of the increased Grashof number is clearly visible, resulting in a stronger asymmetry of the flow caused by aiding and opposing forces acting on the fluid and in steeper temperature gradients at the walls.
While no difference between the strict approximation and the extended formulation can be observed for the streamwise velocity, the influence of the varying thermal diffusivity is clearly visible in the temperature field.
The reference value was chosen as $\kappa_{\mathit{ref}}=\kappa(T_c)$, such that the differences are most pronounced directly at the heated boundary.
Nevertheless, the effects are smaller than the direct impact of the added buoyancy even at these high vapor loads.

Together, these simulations of biperiodic channel flow establish the viability of the chosen numerical approach and the range of applicability for the Boussinesq approximation employed in the governing equations.
They provide a solid starting point for the investigation of the effects of condensation in a comparable geometry.

On this basis, the active scalar approach to simulate flows including phase transition is compared to a subset of the simulations presented by Russo \emph{et al.}\cite{Russo2014}.
Again, a biperiodic channel geometry is considered.
Constant heat flux boundary conditions of $\pm 32\,\mathrm{W/m^2}$ are applied at the walls, such that the net heat flux into the channel is zero.
Flow with bulk Reynolds number $\mathrm{Re}=2333$, corresponding to a friction Reynolds number $\mathrm{Re_{\tau}}=150$, is simulated.
No gravity and consequently no buoyancy are considered.
At the start of the simulation, the fluid temperature is set to $T_0 = 293.15\,\mathrm{K}$, and the vapor concentration to saturation, $c_0 = c_{\mathit{sat}}(T_0)$.
The dimensionless flow parameters are set according to the values for humid air at such conditions, with Prandtl number $\mathrm{Pr}=0.74$ and Schmidt number $\mathrm{Sc}=0.63$.

Instead of an initial droplet distribution, a scalar field representing a continuous reservoir of liquid water dispersed across the whole channel volume is added .
This allows evaporation of liquid if undersaturation conditions are met, mirroring the treatment of condensation discussed previously.
After temperature and vapor concentration reached a statistically steady state, averaging was again performed in time and along the homogeneous directions.
\begin{figure}[htb]
\centering
\includegraphics[scale=1]{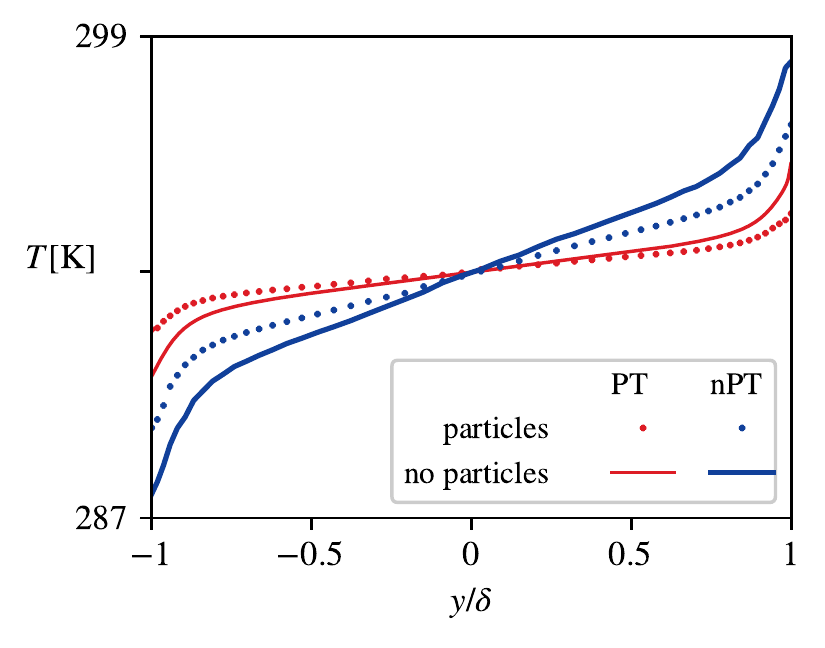}
\caption{Average temperature profiles for four cases. The results from the approach taken in this work, including no particles, but with phase transition (PT), are compared to three cases from Russo \emph{et al.}\cite{Russo2014} covering the remaining combinations.}\label{fig:yT_model_validation}
\end{figure}

Russo \emph{et al.} performed simulations for cases with phase transition (PT) and without phase transition (nPT).
In both cases, particles were included in the investigation, liquid droplets for PT, and inert particles for nPT.
Additionally, a simulation without phase transition and particles was presented as a basis for comparison.

The approach taken in this work completes the possible combinations by including phase transition, but not simulating liquid droplets.

Figure \ref{fig:yT_model_validation} shows the average temperature profiles for the four different cases.
The difference between the case with no particles + nPT and with particles + PT illustrates the modification of the flow due to phase transitions obtained by a two-way coupled multiphase simulation.
These modifications are reflected to a large degree by the temperature profile obtained using the active scalar approach in the case without particles, PT.
The reduction of the temperature difference between the walls due to the latent heat released and absorbed in condensation and evaporation is captured, as well as the reduced temperature gradients.
Discrepancies exist primarily near the walls, where the droplet concentration is high due to turbophoresis\cite{Reeks1983} and the contribution of the droplets to the heat transfer is maximized.

Comparing the active scalar approach to the case with particles + nPT shows that including latent heat and neglecting droplets results a much closer approximation of the full multiphase effects than the converse approach.

Based on the comparison to the two-way coupled multiphase simulation, the simplified approach can be employed for the specific case considered in this study, with parameters reflecting flows at ambient conditions over cool surfaces.
Here, condensation will be primarily constrained to the surface, with condensate adhering to the surface. In this case, the influence of dispersed droplets following the flow is much smaller than in the periodic channel discussed above.
%
%
\section{Investigation setup}
\begin{figure}[htb]
\centering
\includegraphics[width=0.6\linewidth]{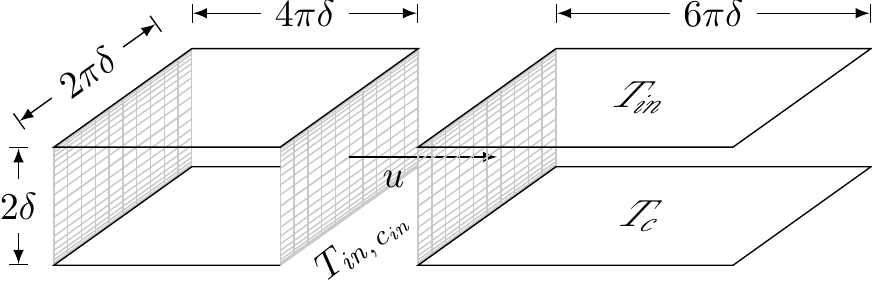}
\caption{Schematic overview of the coupled simulation domains with velocity mapping at the interface and temperature boundary conditions in the primary domain.}\label{fig:domains}
\end{figure}

The governing equations outlined and tested in the preceding sections allow the investigation of flows exposed to temperature gradients causing oversaturation conditions and, consequently, condensation.
A differentially cooled vertical channel geometry is considered in an inlet-outlet configuration to study the impact of condensation on the turbulent flow in such a setup.
The aperiodic setup limits the extent of the thermal boundary layer, ensuring that condensation occurs near the cooled surface and the error due to neglecting droplets in the flow is minimized.
Additionally, the progression of cooling and condensation along the channel length can be investigated in such a geometry.

Values for temperature $T_{\mathit{in}}$ and vapor concentration $c_{\mathit{in}}$ are prescribed at the inlet.
The channel walls are kept at constant temperature, with one wall at $T_h=T_{\mathit{in}}$, while the opposite wall is cooled with respect to the inlet temperature, $T_c<T_{\mathit{in}}$.
By additionally ensuring that the temperature of the cooled wall is below the dewpoint $T_{\mathit{dp}}$ of the fluid at the inlet, which is in turn below the inlet temperature, $T_c<T_{\mathit{dp}}(c_{\mathit{in}})<T_{\mathit{in}}$, a subcooled region can develop within the channel as a consequence of the cooling influence of the wall.
At the opposite wall, setting the temperature boundary to $T_{\mathit{in}}$ prevents the formation of a thermal boundary layer and, consequently, condensation.
Zero-gradient boundary conditions are enforced for the concentration field, letting vapor concentration at the walls evolve in tandem with the field inside the channel.

A precursor simulation is coupled to the system to provide a velocity inlet boundary condition consistent with fully developed turbulent flow\cite{Akselvoll1993,Lund1998,Bellec2017}.
The precursor simulation consists of a separate channel geometry with cyclic boundary conditions in streamwise direction without temperature and concentration fields.
To sustain the flow in the cyclic domain, a self-correcting uniform global pressure gradient is applied, ensuring a constant prescribed volume flux across the interface\cite{Kath2016}.
In addition to feeding back into the inlet of the cyclic domain, the velocity field at the outlet of the precursor simulation is mapped to the inlet of the primary domain.
This effectively creates an infinite-length inflow region where isothermal turbulence can fully develop before being continuously fed into the differentially cooled channel setup.

\begin{figure}[h]
\includegraphics[scale=1]{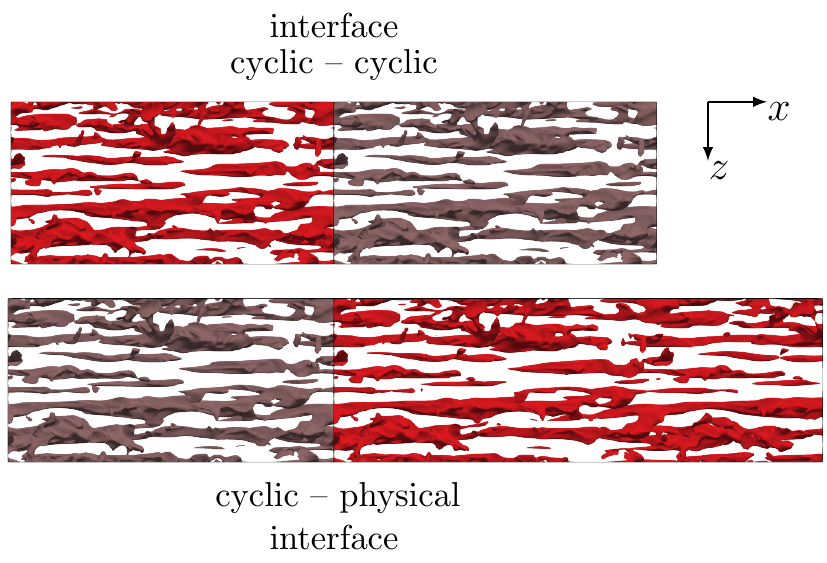}
\caption{Mapping of the velocity field from the precursor outlet back to its inlet (top) and forward towards the primary domain (bottom) visualized via continuous streaks of $u_x^\prime=2u_{\tau}$. The precursor region is replicated again in grey to illustrate the continuity of the streaks.}\label{fig:coupled_u}
\end{figure}

A schematic overview of the simulation domains is shown in Figure \ref{fig:domains}. The resulting flow setup is illustrated in Figure \ref{fig:coupled_u}.
A top view of isosurfaces of the fluctuating streamwise velocity $u_x^\prime=2u_{\tau}$ is drawn in both computational domains.
The top panel shows the continuity of the velocity fields between the cyclic inlet-outlet-planes of the precursor domain, which is repeated for reasons of visualization.
In the bottom panel, the undisturbed transfer of fluid velocity from the precursor to the primary domain is demonstrated in terms of the isosurfaces of the streamwise velocity.

At the channel walls, impermeability and no-slip boundary conditions are applied, while a standard outflow condition is set at the channel outlet.
The pressure boundaries are specified to complement the choices for the velocity field, with a fixed reference pressure set at the outlet and von-Neumann boundary conditions at the inlet.

Flow parameters are chosen to represent a binary mixture of dry air with trace amounts of water vapor, such as expected at standard atmospheric pressure at room temperature.
The diffusive transport of the scalar fields is characterized by the Prandtl number $\mathrm{Pr}=\kappa/\nu = 0.73$ for the temperature and the Schmidt number $\mathrm{Sc}=D/\nu=0.65$ for the vapor concentration reflecting humid air.
Gravity acts along the streamwise direction, causing aiding and opposing buoyant forces to act on the fluid.
Again, the separate contributions due to differences in temperature and vapor concentration can be quantified using the thermal Grashof number $\mathrm{Gr}_T = 38000$ and the solutal Grashof number $\mathrm{Gr}_c=1500$.
Here the characteristic temperature difference $\Delta T = T_{\mathit{in}} - T_c$ refers to the maximal span of temperatures possible within the system, and analogously $\Delta c = c_{\mathit{in}} - c_{\mathit{sat}}(T_c)$ to the corresponding range of concentration.

The bulk Reynolds number is $\mathrm{Re}=1994$, resulting in a friction Reynolds number of
\begin{align}
\mathrm{Re}_{\tau}=\frac{u_{\tau}\delta}{\nu}=135,\label{eq:Reynolds_tau}
\end{align}
where $u_{\tau}$ refers to the friction velocity.
Turbulent flow is expected for channel flow at these values of $\mathrm{Re}$.

The relation between the release of latent heat during condensation and the heating of the system as a consequence is expressed by the Jakob number,
\begin{align}
\mathrm{Ja}=\frac{c_p}{h_v}\Delta T = 0.012.
\end{align}

The combined Richardson number is
\begin{align}
\mathrm{Ri}=\frac{\mathrm{Gr}_T + \mathrm{Gr}_c}{\mathrm{Re}^2} = 0.01.\label{eq:Richardson}
\end{align}
The flow is therefore dominated by forced convection, with only small contributions due to the buoyant forces.
In particular, the Richardson number is much smaller than in the case for which the applicability of the strict Boussinesq approximation was examined ($\mathrm{Ri}=0.07$).
Therefore, using the strict approximation and considering all fluid properties apart from the density as constant is justified going forward.

Further, since the solutal Grashof number is smaller than the thermal Grashof number by an order of magnitude, the direct influence of the solutal buoyancy on the flow is negligible.
On the other hand, the release of latent heat means that a change in thermal buoyancy is leveraged by the concentration change due to condensation.
Given a concentration change $\Delta c^\prime$ due to condensation, the resulting solutal buoyant force is
\begin{align}
F_{\Delta c^\prime}=\beta_c \Delta c^\prime g,\label{eq:force_vapor_diff}
\end{align}
whereas the buoyant force due to the associated release of latent heat is
\begin{align}
\tilde{F}_{\Delta c^\prime}=-\beta_T \frac{h_v}{c_p}\Delta c^\prime g.\label{eq:force_temp_diff}
\end{align}
Given the parameters in this study, the leverage ratio between $\tilde{F}$ and $F$ is
\begin{align}
\frac{\tilde{F}}{F} = -\frac{\beta_T}{\beta_c}\frac{h_v}{c_p}=-22.99.\label{eq:buoyancy_leverage}
\end{align}
Since the condensation mass flux and the velocity field are connected only via the buoyancy term, this leverage ratio shows that the phase transition affects the flow primarily in terms of the associated release of latent heat, far more than by the change in vapor concentration itself.
Note that the negative sign of the leverage ratio signifies that the buoyant forces directly resulting from condensation are opposed to each other, while on the scale of the whole channel, a reduction in temperature leads to a reduction in vapor concentration such that the thermal and solutal buoyancy are aligned.

Two cases are considered in the following.
The primary case includes all fields and interactions as discussed above, in particular the modeling of phase transition (PT).
A reference case omits the vapor concentration field and does not take phase transitions into account (nPT).
By comparing both cases, the influence of condensation can be isolated.

The simulations are initialized using a perturbed laminar channel flow profile\cite{DeVilliers2006}, which is evolved in the absence of the scalar fields until turbulent flow has developed.
In the next phase, the temperature field is included in the simulation, and the system is simulated until the mean temperature within the thermal boundary layer converges.
At this point, the two branches are created.
In the first branch, calculations continue as before and temporal averaging is started to gather data for the reference case nPT, which is used as a baseline for comparison.
For the PT branch, the vapor concentration field and the phase transition model are added to the evolution, and temporal averaging begins after the mean concentration in the vapor boundary layer has converged.
Averages are calculated over intervals of $\Delta t^+=40$ for nPT and $\Delta t^+=36$ for PT.

Discretization with a resolution of $\Delta x^+=5.3$ and $\Delta z^+=2.7$, $\Delta y^+=0.2$ -- $3.3$ is applied consistently to both the primary and the precursor domain to avoid numerical disturbances at the interface.
The resulting mesh consists of $N_x \times N_y \times N_z = (480+320)\times 180 \times 316$ cells.

The upper limit for the discrete time step of the simulation $\Delta t$ is determined by the stability criterion for a second-order central differencing scheme in a system including diffusive scalar transport\cite{Shishkina2004}.
The resulting time step is $\Delta t^+ = 5\cdot 10^{-5}$.
For the combination of low vapor load and moderate subcooling present in the system considered in this study, the integration of the vapor source term derived from the Hertz-Knudsen-Schrage equation given in equation (\ref{eq:HKS_vapor_source}) results in the complete removal of excess vapor in every step of the simulation, since the time scale given by the condensation rate is much smaller than the simulation time step.
This behavior can be mimicked by setting the source term $f$ for equations (\ref{eq:cd_temperature}) and (\ref{eq:cd_concentration}) to
\begin{align}
f(T,c) = \begin{cases} \big( c-c_{\mathit{sat}}(T)\big)/\Delta t,\quad c>c_{\mathit{sat}}(T)\\ \phantom{(}0,\quad \text{else.}\end{cases}\label{eq:simplified_vapor_source}
\end{align}
In this simplified formulation, any vapor in excess of the local saturation value is removed instantaneously, yielding results indistinguishable from the original source term at reduced computational cost.
%
%
\section{Results}
\begin{figure}[htb]
\centering
\includegraphics[scale=1]{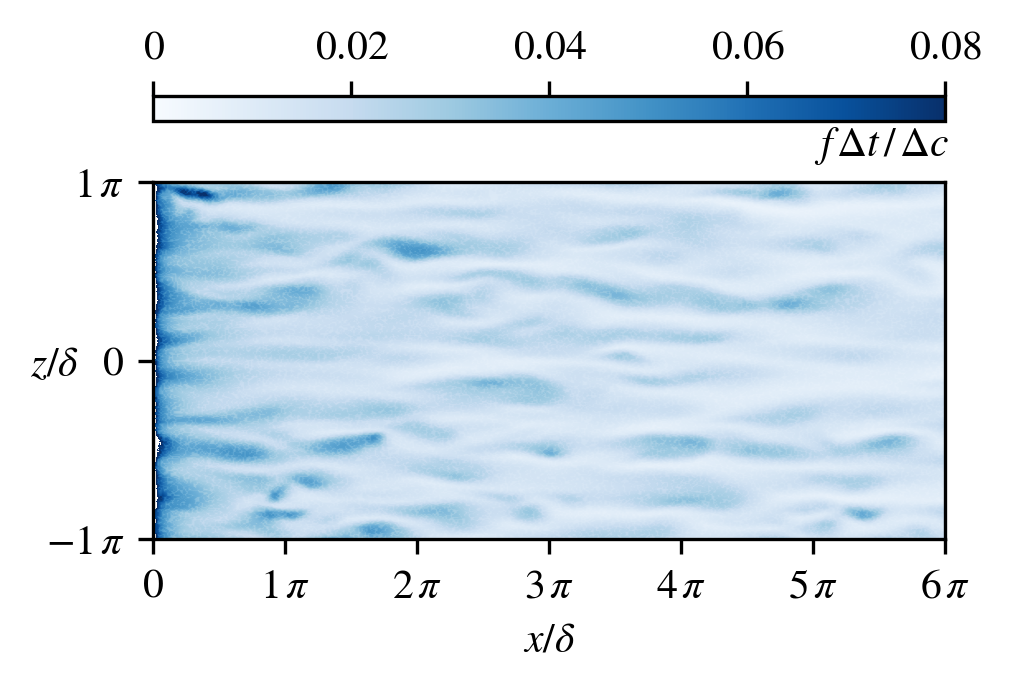}
\caption{Normalized vapor concentration change due to condensation at the cooled wall. High values of $f$ correspond to hot spots of mass transfer across the phase boundary.}\label{fig:source_pattern}
\end{figure}
The DNS of the flow with condensation allows observation of the phase transition rates in the subcooled region within the thermal boundary layer at the cooled wall.
Figure \ref{fig:source_pattern} shows a heatmap of the normalized vapor source term
\begin{align}
\frac{f \Delta t}{\Delta c} = \frac{c-c_{\mathit{sat}}}{\Delta c},
\end{align}
evaluated on the cooled wall boundary patch.
A zone of high mass transfer is visible at $x=0$, where condensation occurs across the full width of the channel as the humid air first meets the cooled wall.
Further downstream, the pattern of condensation is organized into longitudinal, streak-like structures, reminiscent of the wall-layer streaks of fast and slow moving fluid, with a similar spacing of $\lambda_z^+\approx 100$ by visual estimation.\cite{Smith1983}

The inhomogeneity of the condensation pattern and the apparent footprint of coherent flow structures suggest that features of the turbulent flow directly govern the spatial distribution of phase transition events throughout the channel.
Fast-moving fluid originating from the bulk is transported towards the wall in sweeps.
Since it has not yet interacted with the cooled wall, the temperature and vapor concentration will be equal to the values prescribed at the channel inlet, small corrections due to diffusive transport notwithstanding.
At the wall, the fluid is slowed down and cooled by the interaction with the channel boundary, and the vapor concentration drops as a consequence of condensation.
Regions of high condensation rates therefore coincide with the impingement of sweeps onto the cooled surface.
Subsequently, the fluid is ejected from the wall and travels back towards the bulk, moving comparatively slower and carrying less vapor than before.

As a consequence of this vapor transport mechanism, fluid can be classified as pre- and post-interaction based on the instantaneous fluctuations of the scalar fields.
Analogous to equation (\ref{eq:def_theta}), a normalized concentration field is given by
\begin{align}
\xi = \frac{c-c_{\mathit{ref}}}{\Delta c}.\label{eq:def_xi}
\end{align}
The reference values for the given setup are $T_{\mathit{ref}}=T_{\mathit{in}}$ and $c_{\mathit{ref}}=c_{\mathit{in}}$.

Defining the thermal and solutal boundary layer as the set of locations $(x_b, y_b)$ where the cooled wall influences the average flow field due to turbulent mixing or diffusive transport means that $\langle\theta\rangle(x_b,y_b) < \theta_{\mathit{in}}$ and $\langle\xi\rangle(x_b,y_b) < \xi_{\mathit{in}}$ by construction.
Pre-interaction fluid will therefore cause positive fluctuations of the scalar fields, since it is characterized by $\theta_{\mathit{in}}$ and $\xi_{\mathit{in}}$ as outlined above.

Conversely, since this pre-interaction fluid contributes to the overall average, the value of the scalars in fluid that has already interacted with the cooled wall must be below this average value, and therefore cause negative fluctuations.
In combination, these observations yield conditions based on the instantaneous fluctuations of the scalar fields,
\begin{align}
\theta(x,y,z,t) - \langle \theta \rangle (x,y) \begin{cases} > \phantom{-}\epsilon_\theta \quad \text{warm fluid, } \theta_+,\\ < -\epsilon_\theta \quad \text{cold fluid, } \theta_- \end{cases}\label{eq:criterion-theta}
\end{align}
and
\begin{align}
\xi(x,y,z,t) - \langle \xi \rangle (x,y) \begin{cases} > \phantom{-}\epsilon_{\xi} \quad \text{humid fluid, } \xi_+,\\ < -\epsilon_{\xi} \quad \text{dry fluid, } \xi_-. \end{cases}\label{eq:criterion-c}
\end{align}
$\epsilon_{\theta,\,\xi}>0$ can be chosen to improve the signal-to-noise ratio of the criteria, the trade-off being loss of information about fluid closer to the local average than the threshold value.
Here, $\epsilon_{\theta,\,\xi} = 0.01$ are chosen as the limit for the classification.
Averages conditioned in such a way are denoted by $\langle \cdot \rvert \theta_{\pm}\rangle$ and $\langle \cdot \rvert \xi_{\pm}\rangle$, respectively.

Although averaging conditioned on the temperature fluctuations appears to be the natural choice for comparison---as they can be evaluated in both cases irrespective of the presence of water vapor and phase transition effects---it is mandatory to use the fluctuations of the concentration field in the PT case.
Temperature fluctuations cannot be used as the distinguishing criterion here since the injection of latent heat during condensation leads to a momentary increase in temperature, such that post-interaction fluid could possibly be classified as $\theta_+$ and sampled erroneously as pre-interaction.

The conditionally averaged velocity fields are then investigated to confirm the convective transport mechanism leading to the condensation pattern observed in Figure \ref{fig:source_pattern}.

\begin{figure}[hb]
\centering
\includegraphics[scale=1]{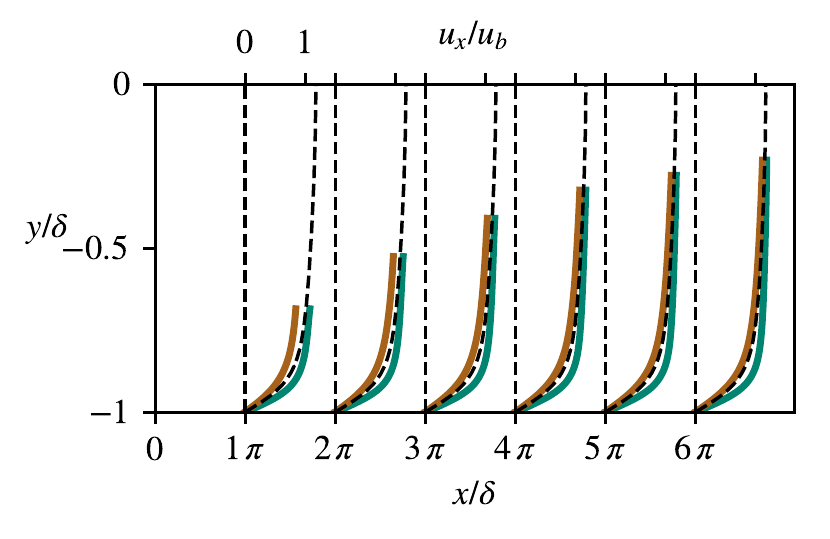}
\caption{Total averaged streamwise velocity profiles {-}{-}{-} $\langle u_x \rangle$ compared to conditionally averaged profiles \textbf{\textcolor{humid}{---}} $\langle u_x \rvert \xi_+\rangle$ and \textbf{\textcolor{dry}{---}} $\langle u_x \rvert \xi_-\rangle$ at different positions along the channel length.}\label{fig:conditional_ux_PT}
\vspace*{2mm}
\centering
\includegraphics[scale=1]{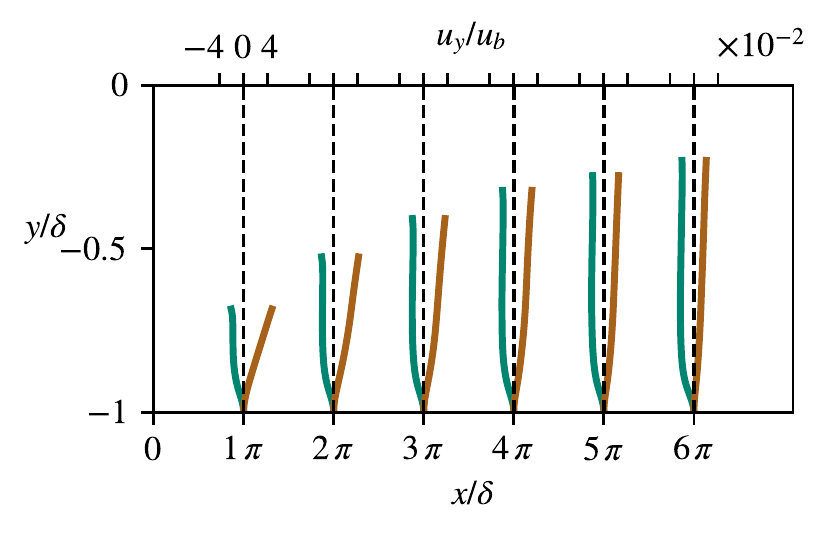}
\caption{Conditionally averaged profiles of the wall-normal velocity \textbf{\textcolor{humid}{---}} $\langle u_y \rvert \xi_+\rangle$ and \textbf{\textcolor{dry}{---}} $\langle u_y \rvert \xi_-\rangle$ at different positions along the channel length.}\label{fig:conditional_uy_PT}
\end{figure}

Figure \ref{fig:conditional_ux_PT} shows profiles of the conditionally averaged streamwise velocity $\langle u_x \rvert \xi_{\pm}\rangle$ compared to the indiscriminately sampled average for PT.
The conditionally averaged values are only well-defined inside the boundary region, since the fluctuations of the scalar fields are zero outside the thermal and solutal boundary layers, resulting in truncated wall-normal velocity profiles.

The ordering $\langle u_x \rvert \xi_-\rangle \leq \langle u_x \rangle \leq \langle u_x \rvert \xi_+\rangle$ can be observed at every position.
Additionally, the wall-normal component of the conditionally averaged velocity $\langle u_y \rvert \xi_{\pm}\rangle$ shows that $\xi_+$-tagged fluid moves towards the cooled wall, while $\xi_-$-tagged fluid exhibits the opposite trend, (Figure \ref{fig:conditional_uy_PT}).

This confirms the hypothesized transport mechanism.
Vapor is carried from the bulk towards the wall in fast-moving sweeps, where it is slowed down and the vapor concentration drops as a consequence of condensation.
After the interaction, the slowed and dried fluid is ejected from the wall and travels back towards the channel center line, causing the growth of the boundary layer as a function of the downstream position.

The evolution of the composition of the boundary layer can be observed in the relation between the sampled averages and the total average of the streamwise velocity.
At the most upstream position shown ($x/\delta=1\pi$), the overall average closely follows the profile of the pre-interaction fluid, suggesting that at this point, the majority of the fluid found in the boundary layer has not yet interacted with the wall.
In contrast, at the downstream positions ($x/\delta=5\pi$ and $6\pi$), a larger portion of the fluid has come into contact with the wall due to the longer residence time.
This is reflected by the overall average tending more towards the profile of $\langle u_x \rvert \xi_-\rangle$.

The analysis of the conditionally averaged velocity profiles illustrates the role of turbulence for the distribution of condensation at the cooled wall.
Convective transport of vapor from the bulk to the wall in turbulent sweeps is the determining factor of phase transition localization, leading to a pattern of condensation that acts as a footprint of the underlying turbulence, as shown in Figure \ref{fig:source_pattern}.

\begin{figure}[t]
\centering
\includegraphics[scale=1]{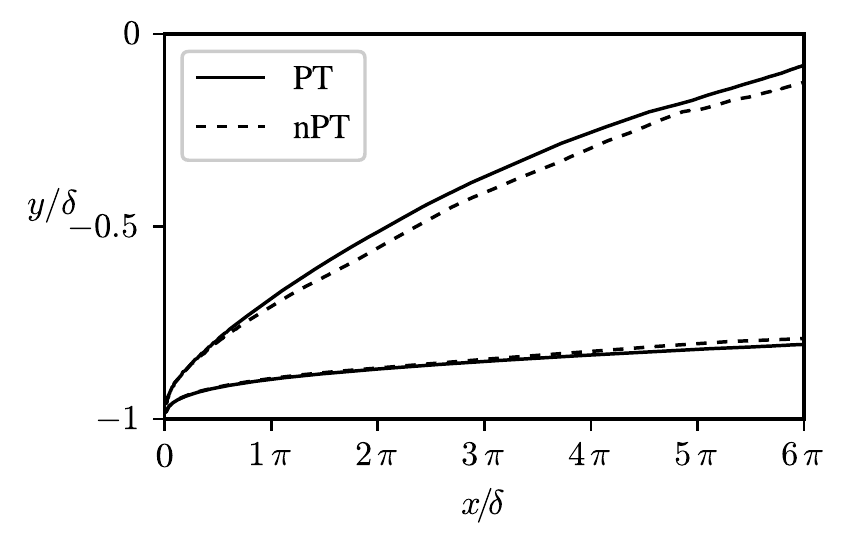}
\caption{$\langle \theta \rangle = 0.75$ and $\langle \theta \rangle = 0.99$ isotherms compared between cases PT and nPT.}\label{fig:isotherms_PT_nPT}
\end{figure}
\begin{figure}[t]
\centering
\includegraphics[scale=1]{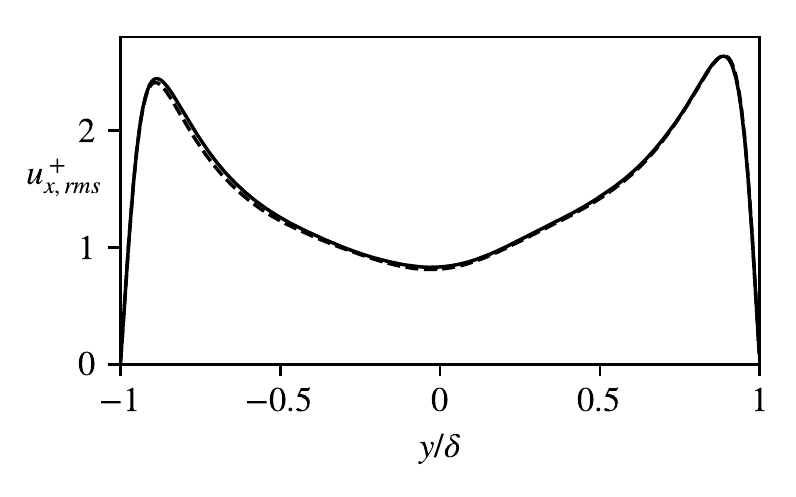}
\caption{Wall-normal profile of the fluctuations of the streamwise velocity component at the channel outlet compared between cases PT and nPT.}\label{fig:ux_rms_PT_nPT}
\end{figure}
The complete coupling implemented for the governing equations causes the influence of the turbulent flow on the occurrence of condensation to feed back to the overall flow.
The injection of heat into the wall-near regions during the phase transition directly affects the thermal boundary layer.
Figure \ref{fig:isotherms_PT_nPT} shows the growth of the thermal boundary layer as characterized by the position of the $\langle \theta \rangle = 0.99$ isotherm along the channel length, compared between cases PT and nPT.
Additionally, the modification of the temperature field in the wall-near region is illustrated by the $\langle \theta \rangle = 0.75$ isotherm.
The influence of the release of latent heat can be observed here.
Fluid with $\langle \theta \rangle<0.75$ extends further into the channel in the case without condensation, indicating that the direct cooling effect of the wall is partially counteracted by the phase transition.
However, the $\langle \theta \rangle = 0.99$ isotherm exhibits the opposite trend, with a thicker overall thermal boundary layer for PT.
This phenomenon is a consequence of the interaction of the localized temperature modification at the wall with the overall turbulent flow field via the buoyant force terms.
The increased average fluid temperature interferes with the damping effect of the aiding buoyant forces on the turbulence.
Figure \ref{fig:ux_rms_PT_nPT} shows the profile of the fluctuations of the streamwise velocity $u_{x\mathit{, rms}}^+$ at the channel outlet.
At this position along the channel, acceleration and deceleration effects of the buoyant forces are more pronounced than at upstream locations, since the residence time of the fluid in the differentially cooled system is maximized.
The well-established damping effect of the aiding buoyant forces acting at the cooled wall\cite{Kasagi1997, Wetzel2019} at $y/\delta=-1$ is visible in the asymmetric profile of the fluctuations, with a reduced peak height near the cooled wall.
Crucially, this damping effect is slightly reduced in the case with condensation, resulting in higher mixing rates and increased transport of cooled fluid towards the bulk, causing the thicker thermal boundary layer observed in Figure \ref{fig:isotherms_PT_nPT}.
This behavior results directly from the negative leverage between solutal and thermal buoyancy during condensation events given in equation (\ref{eq:buoyancy_leverage}).
While the loss in vapor concentration adds to the aiding force acting on the fluid and would therefore increase the damping of turbulence, the leveraged buoyant force caused by the associated release of latent heat results in the opposite behavior, amplifying the turbulence instead.

These modifications of the overall flow field observed in PT together with the connection between the spatial distribution of condensation events and the structure within the turbulent flow suggests that these structures themselves should be sensitive to phase transition in the flow.
To investigate this interaction, spectra of the streamwise velocity fluctuations conditioned on the fluctuations of the scalar fields,
\begin{align}
\langle u_x^\prime \rvert \xi_{\pm} \rangle = \left\langle\left. \sqrt{(u_x - \langle u_x\rangle )^2}\; \right\rvert\xi_{\pm} \right\rangle
\end{align}
and
\begin{align}
\langle u_x^\prime \rvert \theta_{\pm} \rangle = \left\langle\left. \sqrt{(u_x - \langle u_x\rangle )^2}\; \right\rvert\theta_{\pm} \right\rangle,
\end{align}
are extracted for PT and nPT, respectively.
Using the spanwise periodicity of the channel geometry, the energy spectra with respect to the spanwise wavenumber are calculated using fast Fourier transformations.
Since the global effect of buoyancy is small due to the short residence time of the fluid within the system and the limited vapor load and temperature difference (as expressed by the low Richardson number), the streamwise change in the mean velocity is small compared to the magnitude of the fluctuations, such that additional averaging can be performed along this axis.
For better comparability between the different cases, the spectra are normalized with the total energy integrated across all wavenumbers,
\begin{align}
E^\ast(\mathbf{k}) = E(\mathbf{k})\left( \int E(\mathbf{k})\,\mathrm{d}\mathbf{k}\right)^{-1}.
\end{align}

\begin{figure}[t]
\centering
\includegraphics[scale=1]{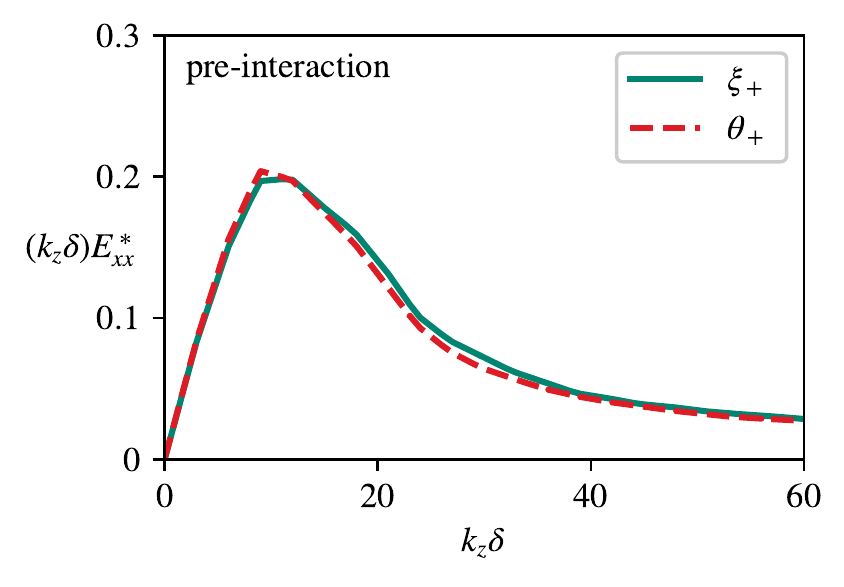}
\caption{Normalized pre-multiplied energy spectrum sampled from pre-interaction fluid \textbf{\textcolor{humid}{---}} $\xi_+$ and \textbf{\textcolor{warm}{---}} $\theta_+$ for PT and nPT.}\label{fig:kzExx_pre_PT_nPT}
\end{figure}

The nondimensional normalized pre-multiplied spectra of the streamwise velocity fluctuations $(k_z \delta)E^\ast_{xx}$ at a distance of $y^+=15$ from the cooled wall, sampled conditionally in accordance with the criteria to distinguish between pre- and post-interaction fluid (\ref{eq:criterion-theta}) and (\ref{eq:criterion-c}), are evaluated in the following.
Figure \ref{fig:kzExx_pre_PT_nPT} shows the spectra of pre-interaction fluid for PT and nPT.
Apart from the negligible effect on the global Grashof number, the vapor concentration field does not directly affect the flow field, only acting indirectly via the leveraged buoyancy during the phase transition.
Consequently, no difference between the two cases can be observed when sampling specifically for fluid that has not yet interacted with the subcooled region near the cooled wall.
In turn, this result underlines that the sampling criteria defined above are well-suited to distinguish between pre- and post-interaction fluid.

\begin{figure}[t]
\centering
\includegraphics[scale=1]{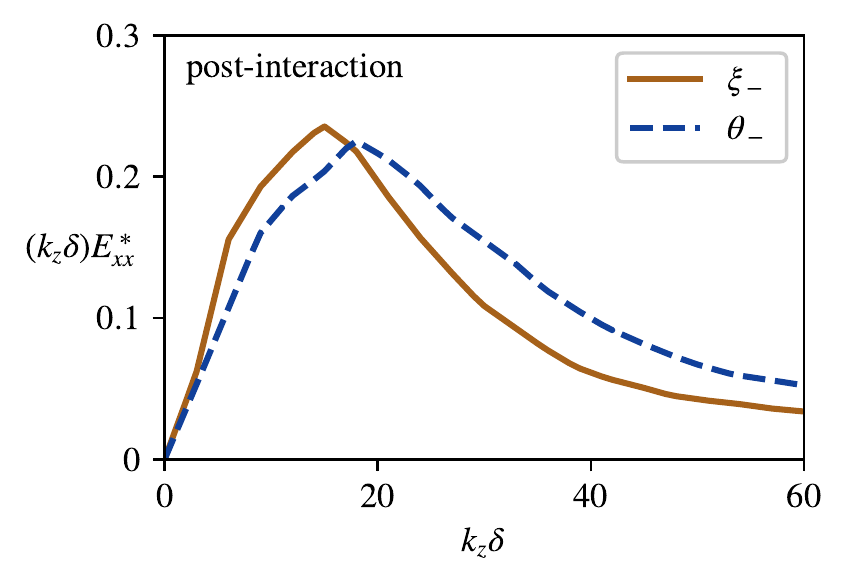}
\caption{Normalized pre-multiplied energy spectrum sampled from post-interaction fluid \textbf{\textcolor{dry}{---}} $\xi_-$ and \textbf{\textcolor{cold}{---}} $\theta_-$ for PT and nPT.}\label{fig:kzExx_post_PT_nPT}
\end{figure}
\begin{figure}[htb]
\centering
\includegraphics[scale=1]{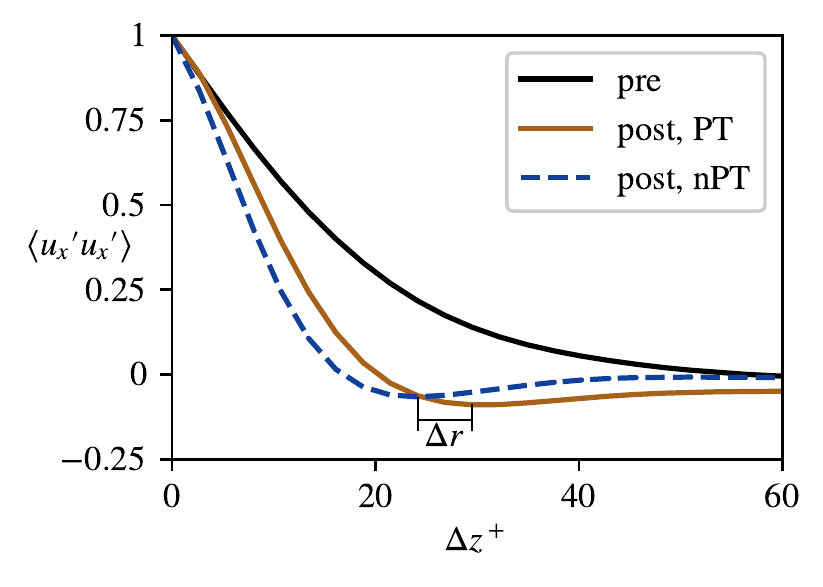}
\caption{Correlation coefficient for the streamwise velocity fluctuations as a function of the spanwise separation $\Delta z^+$ compared between both cases. Only one curve is shown for pre-interaction fluid as the differences between PT and nPT are negligible.}\label{fig:uxux_pre_post_PT_nPZ}
\end{figure}

Figure \ref{fig:kzExx_post_PT_nPT} compares the same quantities after the interaction with the cooled wall.
In both PT and nPT, the spectrum of the fluid ejected from the wall is shifted to higher wavenumbers compared to the sweeps carrying pre-interaction fluid.
This shift is more pronounced in the case without phase transition with regards to both peak position and the high-wavenumber tail of the spectrum.
As with the averaged fluctuations of the streamwise velocity (Figure \ref{fig:ux_rms_PT_nPT}), the release of latent heat at or near the cooled wall during condensation inhibits the cooling of the fluid.
In turn, the turbulent flow structures are affected by the leveraged buoyant forces feeding back from the temperature to the velocity field.
Consequently, the shift towards larger wavenumbers caused by the interaction with the cooled wall without the vapor concentration field is opposed by the effects of the phase transition.
In terms of streak spacing, the difference between PT and nPT is clearly visible in the spanwise spatial correlation coefficient of the velocity fluctuations $\langle u_x^\prime u_x^\prime\rangle$, shown in Figure \ref{fig:uxux_pre_post_PT_nPZ}.
The correlation length $r$, defined as the separation $\Delta z$ from the maximum to the minimum of the correlation coefficient, is increased from $r_{\text{nPT}}^+=24.2$ to $r_{\text{PT}}^+=29.5$ for post-interaction fluid, a change of $\Delta r^+=5.3$.

These results show the impact of the mutual interplay between the release of latent heat during the phase transition and the turbulent flow.
Turbulent flow structures govern the transport of the condensable phase towards the cooled wall and determine the arrangement of the condensation sites.
In turn, the injection of latent heat into the fluid associated with the phase transition changes the turbulent flow structure along the channel by opposing the cooling influence of the channel wall.

Based on this knowledge about the leverage mechanism coupling condensation and the flow fields, the analysis of the behavior for flows with a different ratio of forced to natural convection reduces to the discussion of the effects of thermal buoyancy in such flows\cite{Wetzel2019}.
As $\mathrm{Ri}\to 0$, buoyant forces become negligible compared to the inertia of the fluid, and the effects of the phase transition explored here vanish.
In turn, for flows with a large contribution from natural convection, the opposing effect of the release of latent heat at the cooled wall will grow more important as well.
%
%
\section{Conclusions}
Direct numerical simulations of turbulent convective channel flow including contributions from thermal and solutal buoyancy were performed.
The aspect of latent heat release during phase transitions was included in the simulations by modeling condensation via appropriately chosen source terms for the concentration and temperature fields.
These fields were treated as active scalars, transported through the system by convection and diffusion.
This approach allowed the investigation of this aspect of phase transitions without a full multiphase simulation.

The setup of the system with an inlet-outlet simulation domain allowed for the analysis of the evolution of the flow as a function of the streamwise coordinate, showing that the residence time of the fluid within the region of influence of the cooled wall determines the degree to which phase transition effects affect the fluid on average.
Additionally, this configuration lends itself more readily to comparisons with experimental investigations of similar systems in the future.

Simulations of an identical system without the inclusion of the vapor field and phase transition modeling provided a baseline for comparisons against which the effect of the condensation on the flow could be evaluated.

The analysis of the mutual interplay between the turbulent flow and condensation revealed a clear bidirectional link between both aspects of the system.
The arrangement of hot spots of mass transfer across the phase boundary at the cooled wall reflects the structure of the alternating longitudinal streaks of high- and low-velocity fluid in the turbulent flow.
This condensation pattern results from the mechanism of convective vapor transport.
Fast-moving sweeps carry fluid with high vapor concentration towards the wall, thus creating oversaturated conditions and causing condensation events.

In turn, the turbulent flow itself is modified by the effects of the phase transition due to the coupling via the buoyant forces.
Here, the influence of buoyancy induced by the release of latent heat is opposed and far greater than the solutal contribution from the underlying change in vapor concentration during condensation.

Collectively, the modifications of the turbulent flow are characterized by the opposition to the cooling effect caused by the injection of heat into the subcooled region near the wall.
While the system is very efficient in removing the added thermal energy, such that an increase in average temperature is found only directly at the wall, the high spatial and temporal correlation between the associated generation of temperature spikes during condensation events and the underlying turbulent flow structure reduces the damping effect of the aiding buoyant force acting on the fluid.
This results in a slightly elevated peak turbulence intensity compared to the case without phase transition.
The small effect size is a consequence of the small overall importance of buoyancy compared to the inertia of the forced convection for the flow parameters investigated here.

Conditional sampling based on the instantaneous fluctuations of the scalar fields proves to be a valuable tool for distinguishing between pre- and post-interaction fluid.
Applied to the pre-multiplied spectra of the streamwise velocity fluctuations, they again show that the injection of latent heat during condensation serves to resist the changes exhibited by the system without phase transition, as the shift towards higher wavenumbers is considerably reduced.
This reflects a significant modification of the spanwise spacing of post-interaction streaks.

The treatment of condensation is limited to the effect of the release of latent heat.
In particular, the drop of the partial pressure of the vapor due to the phase transition\cite{Bukhvostova2014}, inducing compressibility effects, is not considered here.
Additionally, the deposition of condensate at the walls and the subsequent formation of droplets provides another mechanism for interaction with the turbulent flow.

Including these aspects of condensation is possible while at the same time preserving the single-phase approach, promising a simplified, yet accurate method of simulating flow with phase transition for systems similar to those presented in this investigation.
%
%
%
\section*{Acknowledgments}
The authors thank Annika K\"ohne for proofreading the paper.
\bibliographystyle{aipnum4-1}
\bibliography{Revised1-Condensation-induced_flow_structure_modifications_in_turbulent_channel_flow_investigated_in_DNSNotes.bbl}
\end{document}